\documentclass[%
 reprint,
 amsmath,amssymb,
 aps,
floatfix,
]{revtex4-2}

\usepackage{color}
\usepackage{graphicx}
\usepackage{dcolumn}
\usepackage{bm}

\begin{document}

\preprint{APS/123-QED}

\title{Real surreal trajectories in pilot-wave hydrodynamics}

\author{Valeri Frumkin}
\email{valerafr@mit.edu}
\affiliation{Department of Mathematics, Massachusetts Institute of Technology.}

\author{David Darrow}
\email{ddarrow@mit.edu}
\affiliation{Department of Mathematics, Massachusetts Institute of Technology.} 
 
\author{Ward Struyve}
\email{ward.struyve@kuleuven.be}
\affiliation{Instituut voor Theoretische Fysica \& Centrum voor Logica en Filosofie van de Wetenschappen, KU Leuven}

\author{John W. M. Bush}
 \email{bush@math.mit.edu}
\affiliation{Department of Mathematics, Massachusetts Institute of Technology.}

\begin{abstract}
In certain instances, the particle paths predicted by Bohmian mechanics are thought to be at odds with classical intuition. A striking illustration arises in the interference experiments envisaged by Englert, Scully, S\"ussmann and Walther, which lead the authors to claim that the Bohmian trajectories can not be real and so must be `surreal'. Through a combined experimental and numerical study, we here demonstrate that individual trajectories in the hydrodynamic pilot-wave system exhibit the key features of their surreal Bohmian counterparts. These real surreal classical trajectories are rationalized in terms of the system's non-Markovian pilot-wave dynamics. Our study thus makes clear that the designation of Bohmian trajectories as surreal is based on misconceptions concerning the limitations of classical dynamics and a lack of familiarity with pilot-wave hydrodynamics.

\end{abstract}

\maketitle

\paragraph*{Introduction:} 

The standard theoretical formulation of quantum mechanics does not describe particle paths. Bohmian mechanics~\cite{BohmHiley,HollandBook,DurrTeufel} 
provides such a description by asserting that
particles move under the influence of the wave function, with a velocity equal to the quantum velocity of probability in the standard quantum formalism. 
Specifically, the dynamics of a single particle with position ${\bf X}$ is prescribed by the {\it guidance equation},
\begin{equation}\label{guidance}
    \frac{d {\bf X}(t)}{dt}=\frac{\hbar}{m}\textrm{Im}\!\left[\!\frac{\pmb{\nabla}\psi\big({\bf X}(t),t\big)}{\psi\big({\bf X}(t),t\big)}\!\right]\!,
\end{equation}
where $\psi$ is the wave function satisfying the Schr\"odinger equation with a potential $V$. Taking the time derivative of (1) yields a Newtonian-type trajectory equation 
\begin{equation}
m  \frac{d^2 {\bf X}(t)}{dt^2} = -{\boldsymbol{\nabla}} V({\bf X}(t)) -{\boldsymbol{\nabla}} Q({\bf X}(t),t) ,
\end{equation}
where $\smash{Q=- \frac{\hbar^2}{2m} \frac{1}{|\psi|} \nabla^2 |\psi|}$, the quantum potential, engenders the departure from classical mechanics.
The statistical predictions of Bohmian mechanics are in accord with those of the standard quantum theory~\cite{BohmHiley,HollandBook,DurrTeufel}.

Among the successes of Bohmian mechanics is a self-consistent 
description of single-particle diffraction and interference. 
The wave function guides the particles through the slits according to Eq.~\eqref{guidance}, eventually leading to the familiar diffraction and interference patterns building up over time. The trajectories for the canonical double slit experiment were first computed by Philippidis {\em et al.}\ \cite{Philippidis1979}, who demonstrated that the emergent diffraction pattern predicted by Bohmian mechanics is consistent with that observed provided one chooses an ensemble of initial positions prescribed by 
a Gaussian distribution of impact parameters at the slits.
While direct measurement would alter the wave function and hence the particle trajectory, the averaged trajectories revealed by weak measurement strongly resemble the trajectories predicted by Bohmian mechanics~\cite{Kocsis2011}.

In 1992, Englert, Scully, S\"ussman, and Walther (ESSW)~\cite{englert_surrealistic_1992} proposed an interference experiment intended to expose the shortcomings of Bohmian mechanics. They claimed that ``the Bohm trajectory is here macroscopically at variance with the actual, that is: observed, track. Tersely: Bohm trajectories are not realistic, they are surrealistic''. Such criticisms have been countered by several authors~\cite{dewdney_how_1993,durr_comment_1993,barrett2000,hiley_2000,Laloe}, on the grounds that the type of observation envisioned by ESSW cannot be expected to reliably reveal the Bohmian trajectory. Nevertheless, ESSW~\cite{Englert1993} maintained that ``one cannot attribute reality to the Bohm trajectories'' \cite{Englert1993}, ``particles do not follow the Bohm trajectories as we would expect from a classical type model'' \cite{Scully_1998} and finally declared ``the Bohmian picture to be at variance with common sense'' \cite{Scully_1998}. In the absence of experimental measurements of actual particle paths, the designation of Bohmian trajectories as real or surreal necessarily depends only on one's preconceptions at to how quantum particles should behave. 
In their discussion of ESSW, Aharonov and Vaidman~\cite{Aharonov1996} conclude that ``The examples considered in this work do not show that Bohm's causal interpretation is inconsistent. It shows that Bohmian trajectories behave not as we would expect from a classical type model''. 
The aim of this letter is to present a purely classical wave-particle system that exhibits surreal trajectories. In so doing, we demonstrate that surreal trajectories are not at odds with classical intuition informed by a familiarity with pilot-wave hydrodynamics.

\begin{figure}[t!]
\noindent \begin{centering}
\includegraphics[width=20.3pc]{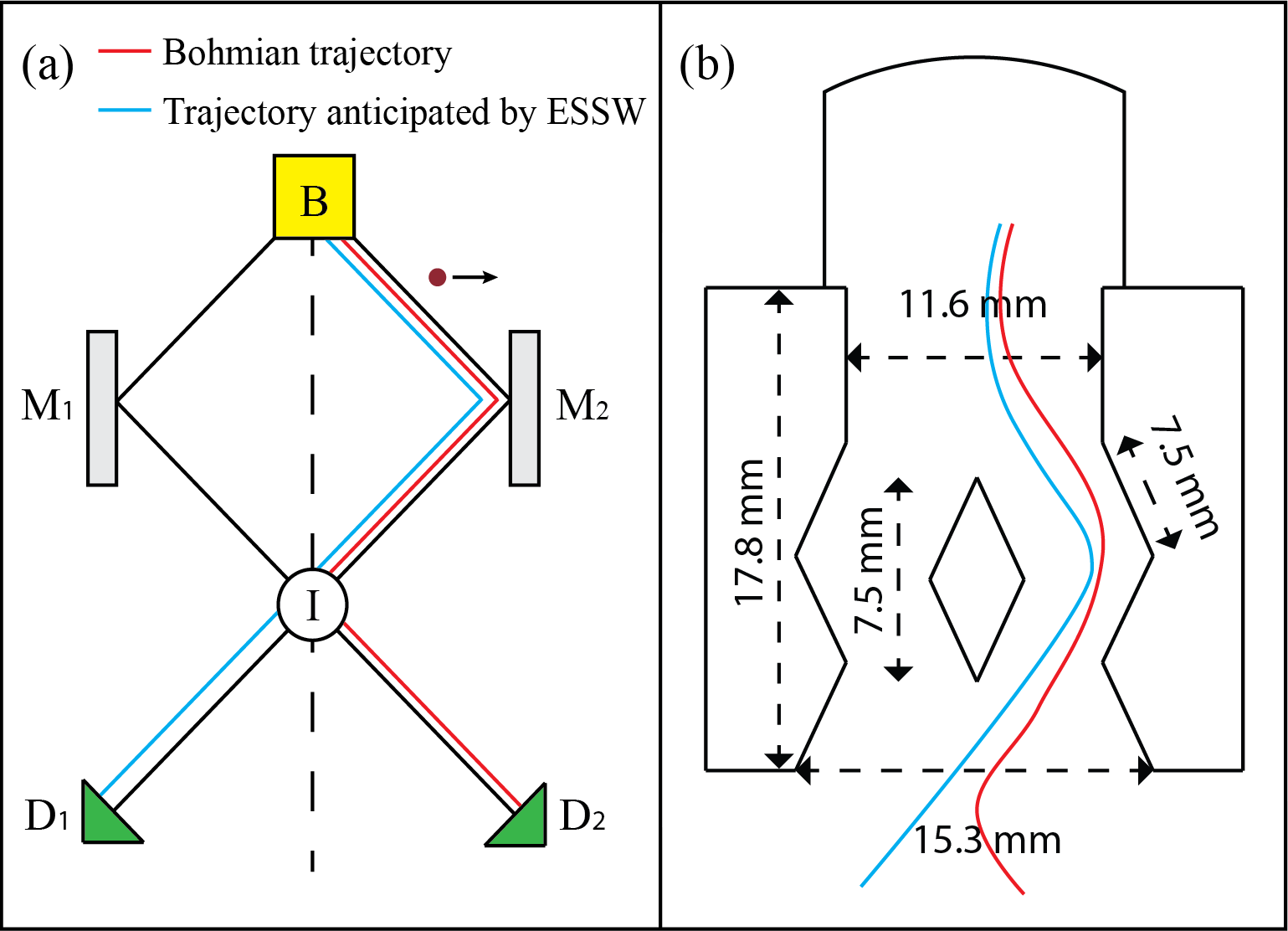}
\par\end{centering}
    \caption{(a) A variant of the interferometer setup considered by ESSW \cite{englert_surrealistic_1992}. An incoming wave packet is split by a beam splitter $B$ and reflected by the mirrors $M_1$ and $M_2$. The wave packets interfere in the region $I$ and then move towards the detectors $D_1$ and $D_2$. The blue path represents the particle trajectory anticipated by ESSW, while the red path is that predicted by Bohmian mechanics. The red dot represent the one-bit, which-way detector. (b) Our experimental arrangement. A walking droplet is released from a launching area towards a beam-splitter that directs the droplet towards one of two submerged barriers with equal probability. The blue and red paths are the possible analogues of those in the ESSW setup.}  
\end{figure}

\paragraph*{The ESSW setup:} The arrangement considered by ESSW is depicted in Fig.~1(a). A single particle is directed towards a beam splitter $B$. The associated wave packet $\psi$ is thus represented by the superposition, $\psi_+ + \psi_-$, of packets $\psi_+$ and $\psi_-$ that propagate, respectively, towards the mirrors $M_1$ and $M_2$. Note the perfect symmetry about the system's vertical centerline. After reflection from the mirrors, the packets cross before eventually reaching the detectors $D_1$ and $D_2$. According to Bohmian mechanics, the particle will initially go either left or right, with
equal probability. In Fig.~1(a), the particle initially goes right, bounces off the mirror $M_2$ and then enters the interference region $I$. However, the Bohmian particle does not continue its path towards $D_2$ but instead turns around in region $I$, then proceeds toward $D_1$. 
This behavior is generic: no Bohmian trajectories cross the symmetry axis; instead, they turn around in the region $I$. Such behavior is at odds with the behavior anticipated by ESSW, who presumed that the particles should not change direction at $I$. If the mirror $M_2$ were to be removed or a detector were to be placed in the path towards $M_2$, there would be no interference between $\psi_+$ and $\psi_-$ and the Bohmian path would correspond to that anticipated by ESSW.

ESSW also assumed that a one-bit which-way detector along one of the paths would reveal the path taken by the particle {\it after} the particle passes through the interferometer. They thus hoped to show that the Bohmian trajectory is at odds with the inferred trajectory. 
A variant of the ESSW setup using double-slits was proposed by Braverman and Simon \cite{braverman_proposal_2013} and realized experimentally by Mahler {\it et al.}~\cite{mahler_experimental_2016}, who reported that the trajectories inferred from weak measurement are consistent with the predictions of Bohmian mechanics. 
We here demonstrate that the surreal character of the Bohmian trajectories also arises in pilot-wave hydrodynamics.

\paragraph*{Pilot-wave hydrodynamics:} In 2005 Yves Couder and Emmanuel Fort discovered that a millimetric droplet may self-propel along the surface of a vibrating liquid bath by virtue of a resonant interaction with its own quasi-monochromatic wave field \cite{couder_walking_2005}. In a seminal 2006 paper, Couder and Fort used this system to demonstrate single- and double-slit diffraction and single-particle interference \citep{couder_single-particle_2006}, phenomena previously thought to be exclusive to the microscopic quantum realm. Their study has been revisited several times, the diffraction and interference effects confirmed~\cite{Pucci2018,Ellegaard2020}. This hydrodynamic pilot-wave system~\cite{Bush2015a} has since been used to establish many other hydrodynamic quantum analogs~\cite{bush_hydrodynamic_2020}, including unpredictable tunneling through potential barriers \cite{eddi_unpredictable_2009}, quantization of orbital states \cite{fort_path-memory_2010}, the emergence of wave-like statistics in corrals \cite{harris_wavelike_2013, saenz_statistical_2018}, Friedel oscillations \cite{saenz_hydrodynamic_2020}, and hydrodynamic spin lattices \cite{saenz_emergent_2021}.

A key feature of the hydrodynamic pilot-wave system is that the droplet dynamics is non-Markovian~\cite{Bush2015a}. Specifically, the propulsive force on the droplet is prescribed by the local slope of the wave field, the form of which necessarily depends on the droplet's history. The wave field persists for a time prescribed by the bath's vibrational  acceleration, $\gamma$. The bath thus effectively serves as the `memory' of the droplet~\cite{Eddi2011a}, and the system memory is prescribed by the proximity of $\gamma$ to the Faraday threshold, $\gamma_F$, the critical vibrational acceleration at which Faraday waves form on the bath surface in the absence of the drop.
Notably, the quantum features arise exclusively in the high-memory limit, $\gamma \rightarrow \gamma_F^{-}$, when the effects of the pilot wave and the system memory are most pronounced.

In several instances, non-local features of the walking droplet system might be misinferred if the influence of the pilot-wave is not given due consideration. For example, submerged pillars and posts give rise to long-range lift forces on walking droplets that are mediated by the pilot wave~\cite{Harris2018,saenz_hydrodynamic_2020}. In the walker double-slit experiment~\cite{couder_single-particle_2006}, the presence of the second slit was seen to have an influence on droplets passing through the first~\cite{Pucci2018,Ellegaard2020}. Without due consideration  
of the dynamical significance of the pilot-wave field and the itinerant non-Markovian droplet dynamics, such wave-mediated forces appear to act at a distance and so to be spatially nonlocal. Moreover, Bush and Oza~\cite{bush_hydrodynamic_2020} discuss settings in which the mean-pilot-wave potential~\cite{Durey2018} plays a role akin to the quantum potential in Bohmian mechanics. 

\begin{figure*}[t!]
\includegraphics[width=42pc]{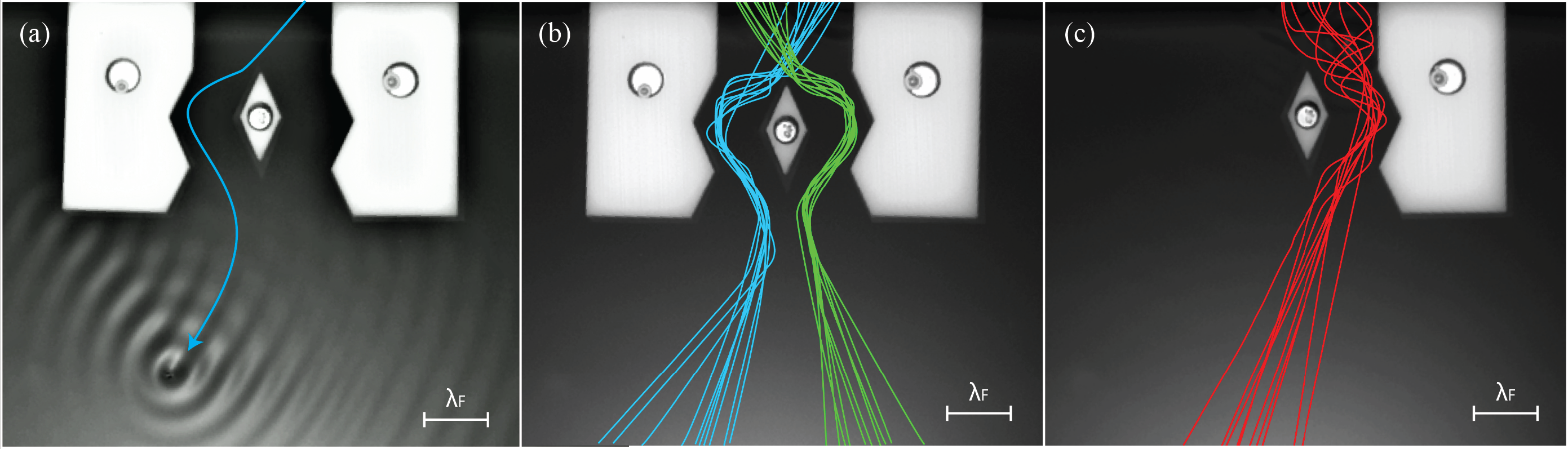}
\caption{\label{fig2} Droplet trajectories in the hydrodynamic pilot-wave system. (a) A single particle trajectory, along with the pilot wave field at the instant that the drop is at the position shown. (b) In a symmetric setup, the droplet enters the right or left channel with equal probability, after which it is deflected away from the system centerline, resulting in a `surreal' trajectory. 
Twenty such trajectories are shown. (c) When one of the barriers is removed, the symmetry of the system is broken. The walking droplet is then reflected away from the remaining barrier, resulting in the trajectory that one might expect.  }
\end{figure*}

Walking droplets exhibit non-specular reflection from submerged planar barriers \cite{pucci_non-specular_2016}. The weak dependence of the reflection angles on the angle of incidence allows one to use submerged planar barriers as reflectors, or analog mirrors.
We proceed by presenting a hydrodynamic analog of the ESSW thought experiment using submerged planar barriers as reflectors. In so doing, we demonstrate that surreal trajectories may arise in the hydrodynamic pilot-wave system.

\paragraph*{Experiments:} 
Our experimental system consists of a circular bath filled with a $7.0 \pm 0.3$ mm deep layer of silicon oil with surface tension $\sigma=0.0209$ N/m, viscosity $\nu=20$ cSt, and density $\rho=0.965\times 10^{-3}\,\,\,\text{kg}/\text{m}^{3}$.
The system is vibrated vertically by an electromagnetic shaker with forcing $F(t)=\gamma\cos(2\pi ft)$, with $\gamma=3.79$ g and $f=75$ Hz being the peak vibrational acceleration and frequency, respectively. In all experiments reported here, the Faraday threshold, at which the flat free surface destabilizes to a pattern of sub-harmonic Faraday waves with wavelength $\lambda_F = 5.6$ mm, was $\gamma_F=3.82$ g. 
Spatial uniformity of the bath vibration was insured by connecting the shaker to the bath with a steel rod coupled to a linear air bearing~\citep{harris_generating_2015}. The vibrational forcing was monitored with two accelerometers placed on opposite sides of the bath, insuring a constant vibrational acceleration amplitude to within $\pm 0.002$ g.
The droplet trajectories and their guiding wave field were captured through a semi-reflective mirror that was positioned at 45° between the bath and a charge-coupled device (CCD) camera that was mounted directly above the setup. The bath was illuminated with a diffuse-light lamp facing the mirror horizontally, yielding images with bright regions corresponding to horizontal parts of the surface, specifically extrema or saddle points (See Fig.~2(a)). 

The topographical configuration used in our experiments is depicted in Fig.~1(b). A walking droplet is confined to a launching pad, where it wanders in an irregular fashion until being ejected towards a submerged rhombus that forces the droplet towards one of two submerged barriers with equal probability.
The launch randomizes the droplet's initial conditions, while the rhombus serves as a beam splitter, and the submerged barriers act as reflectors. In the first experiment, the droplet passes the beam splitter and is reflected away from the adjacent barrier; subsequently, it changes direction again before reaching the centerline.
Fig.~2(b) shows $20$ such trajectories (10 from each side), obtained in a continuous experiment with a single droplet. 
If the influence of the pilot-wave is not duely considered, one could only rationalize the second change in direction in terms an external force. 
In reality, this apparently non-local behavior is a manifestation of the local influence of the pilot-wave that interacts with both barriers and the droplet in such a way as to produce the surreal trajectories depicted in Fig.~2(b). 
In the second experiment, one of the barriers is removed while the rest of the setup remains unaltered (Fig.2(c)). With this asymmetric configuration, after the droplet is reflected from the barrier, it does not change direction appreciably. The second experiment underscores the importance of the presence of both barriers for surreal droplet trajectories.
A video of the two experiments is available in the supplementary materials (SV1).

\paragraph*{Numerical simulations:} 
We proceed by adopting the numerical model of Faria~\cite{faria_model_2017}, as has been shown to provide a robust description of walking droplet-boundary interactions in a number of settings, including reflection from walls~\cite{pucci_non-specular_2016}, scattering from a submerged circular well~\cite{saenz_emergent_2021} and diffraction through slits~\cite{Pucci2018}. The model synthesizes the walker wave model of Milewski {\it et al.}~\cite{milewski_faraday_2015} with the trajectory equation of Molacek and Bush~\cite{molacek_drops_2013}. The effect of the topography on the waves is captured through encorporating its influence on the
local phase speed of the pilot wave. 
The distinct advantage of the simulations is that they enable the precise prescription of the initial conditions for the droplet trajectories, computation of the associated wave fields and characterization of the influence of increasing memory on the droplet dynamics.

\begin{figure}[t!]
\includegraphics[width=20.5pc]{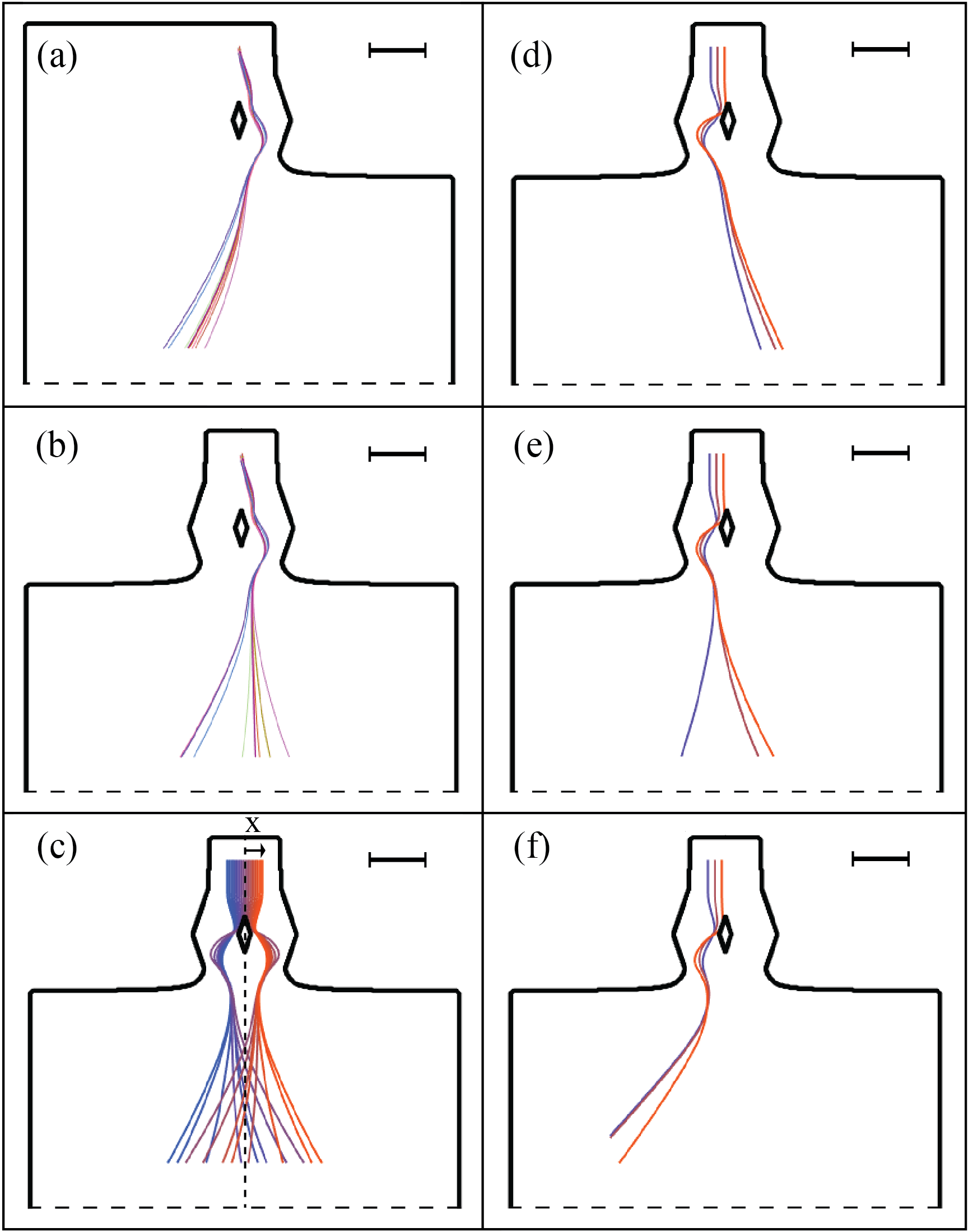}
\caption{\label{fig3} Simulated droplet trajectories in the hydrodynamic pilot-wave system. In (a)-(c), the vibrational acceleration $\gamma = 0.905\gamma_F$. (a) When only one reflector is present, the droplet follows the expected trajectory, regardless of the initial conditions. (b) In the presence of a second reflector, depending on the initial conditions and the system memory, the droplet may follow a `surreal' trajectory, and so never cross the centerline.  (c) An ensemble of initially vertical trajectories with different values of the impact parameter $x$. (d)-(f) The dependence on memory of a trio of originally vertical trajectories. As the memory parameter is increased from (d) $0.88\Gamma_F$ to (e) $0.9\Gamma_F$ to (f) $0.92\Gamma_F$, all trajectories transform from expected to surreal. Scale bar: $5\lambda_F$.}
\end{figure}

The results of the numerical simulations are presented in Fig.~3. In the absence of a second reflector (Fig.~3(a)), the droplets follow the expected trajectories, being reflected by the barrier on the right, then proceeding to cross the centerline. When a barrier on the left is added (Fig.~3(b)), the resulting trajectories may exhibit surreal behaviour, depending on the initial conditions and system memory. Fig.~3(c) illustrates the dependence of the droplet trajectory on the impact parameter ($x$, the distance from the centerline) for an ensemble of drops launched downwards.
Fig.~3(d)-(f) demonstrate that the surreal nature of the droplet trajectories becomes more pronounced as the system memory is increased. At sufficiently high memory, the pilot wave excited by the droplets extends 
to the outer boundaries of the domain.
Thus, our simulations were limited to a memory value of $0.92 \gamma_F$, above which the droplets were affected by the domain's outer edges.

Fig.~4(a) shows the mean wave field that results from calculating the weighted average of the pilot wave forms arising over all droplet trajectories. 
Fig.~4(b) depicts the trajectories used in this ensemble, which were weighted according to a Gaussian distribution in impact parameter in order to compute the mean wave field.
The relation between this mean wave field and the quantum potential
in Bohmian mechanics will be the subject of future investigation. \\

\paragraph*{Discussion:} 

The walking-droplet system has been shown to exhibit several features previously thought to be exclusive to the quantum realm~\cite{bush_hydrodynamic_2020}. We have shown here that the surreal trajectories predicted by Bohmian mechanics are another such feature.
The surreal character of both Bohmian and droplet trajectories may be attributed to wave-mediated forces. 
However, a number of notable distinctions should be made between the trajectories arising in Bohmian mechanics and pilot-wave hydrodynamics.
First, the Bohmian trajectories  cannot be measured precisely without being disturbed. 
Second, according to Bohmian mechanics, particles are guided by the wave function $\Psi$, whose form is uninfluenced by the particle.
Conversely, the trajectories in pilot-wave hydrodynamics are directly observable and result from the droplet's interaction with its own  wave field. Specifically, the droplet navigates its pilot-wave field, a local potential of its own making.
One might thus attribute the designation of `surreal' to Bohmian trajectories to ESSW's lack of familiarity with the walking droplet system, wherein such trajectories may be rationalized in terms of non-Markovian, classical, pilot-wave dynamics. 

\begin{figure}[t!]
\includegraphics[width=20.4pc]{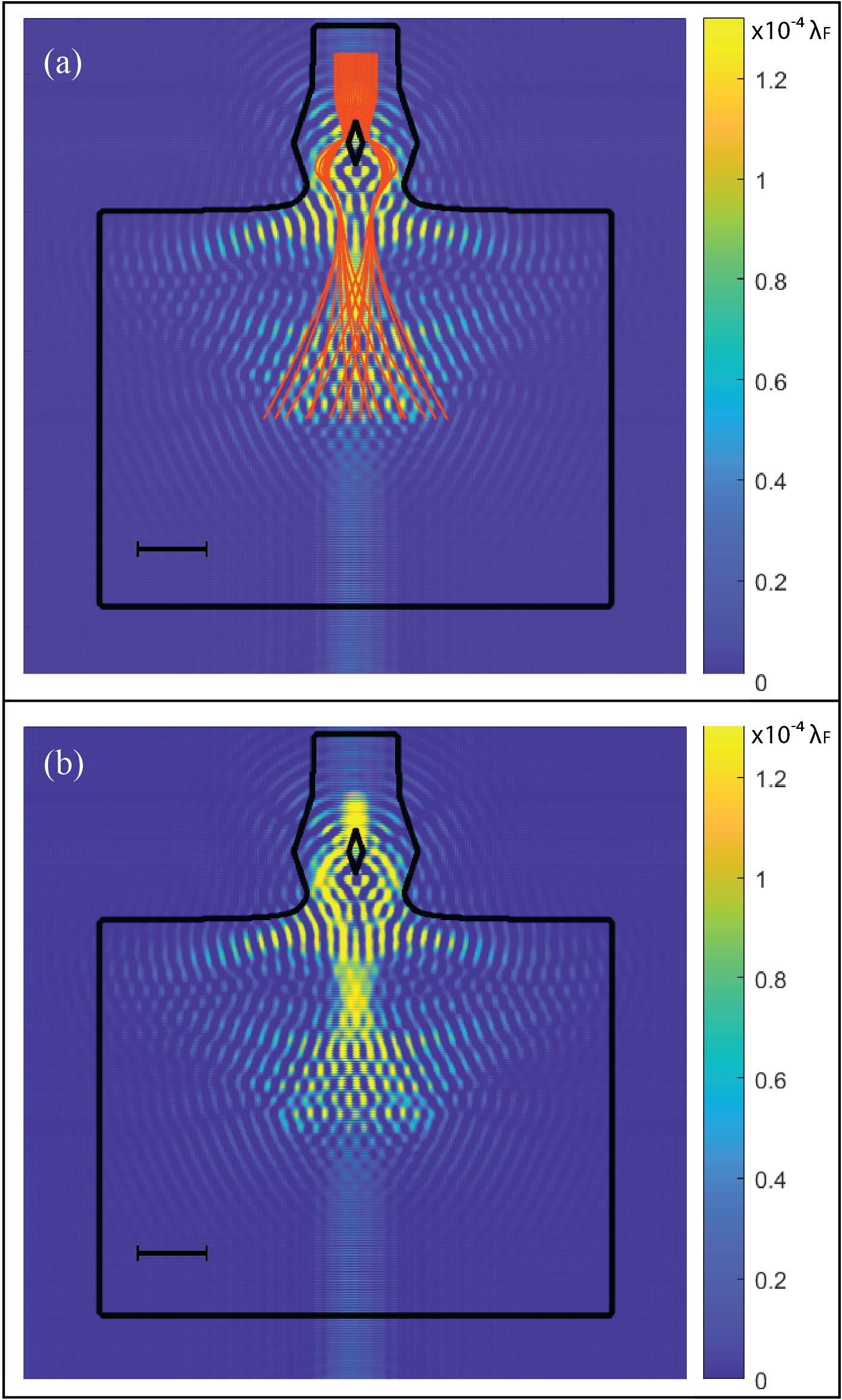}
\caption{\label{fig4} The mean wave-field generated by averaging simulated droplet trajectories with a Gaussian distribution of initial impact parameters, with a standard deviation of $1.4 \lambda_F$, (a) with and (b) without the associated droplet trajectories overlaid. 
The wave amplitude is presented in units of the system's Faraday wavelength. Scale bar: $5\lambda_F$.}
\end{figure}

Prior work~\cite{Bush2015a,bush_hydrodynamic_2020} has shown  that the walking droplet system is closer in form to de Broglie's original double-solution pilot-wave theory~\cite{colin2017}, according to which a quantum particle has an internal vibration at the Compton frequency that generates its own guiding wave, and the resulting pilot-wave dynamics gives rise to emergent statistics described by the standard quantum theory. 
Our study provides further motivation for the revisitation and extension of de Broglie's double-solution program, informed by the walking-droplet system~\cite{borghesi_equivalent_2017,dagan_hydrodynamic_2020,durey_hydrodynamic_2020,drezet_mechanical_2020}.

\vspace*{0.1in}
The authors gratefully acknowledge the financial support of the United States Department of State (V.F.), the National Science Foundation grant CMMI-2154151 (J.B.), and the Research Foundation Flanders (Fonds Wetenschappelijk Onderzoek, FWO), Grant No.\ G066918N (W.S.). We thank Boris Filoux and Nicolas Vandewalle for valuable discussions.

\bibliographystyle{ieeetr}
\bibliography{ST.bib}

\end{document}